\documentclass[twocolumn,showpacs,preprintnumbers]{revtex4}
\usepackage{graphicx}
\usepackage{dcolumn}
\usepackage{bm}

\begin{document}

\newcommand{\homno} {HoMnO$_{3}$}
\newcommand{\highTphase} {$P6'_{3}c'm$}
\newcommand{\midTphase} {$P6'_{3}cm'$}
\newcommand{\lowTphase} {$P6_{3}cm$}

\preprint{APS/123-QED}
\title{Magnetic Order and Spin Dynamics in Ferroelectric HoMnO$_{3}$}
\author{O. P. Vajk,$^{1}$ M. Kenzelmann,$^{1,2}$ J. W. Lynn,$^{1}$
  S. B. Kim,$^{3}$ and S.-W. Cheong$^{3}$}
\affiliation{$^{1}$NIST Center for Neutron Research, Gaithersburg, MD 20899\\
  $^{2}$Department of Physics and Astronomy, Johns Hopkins
  University, Baltimore, MD 21218 \\
  $^{3}$Department of Physics and Astronomy, Rutgers University,
  Piscataway, New Jersey 08854}
\date{\today }

\begin{abstract}
Hexagonal HoMnO$_{3}$ is a frustrated antiferromagnet (T$_{N}$=72 K)
ferroelectric (T$_{C}$=875 K) in which these two order parameters are
coupled. Our neutron measurements of the spin wave dispersion for the
$S=2$ Mn$^{3+}$ on the layered triangular lattice are well described by a
two-dimensional nearest-neighbor Heisenberg exchange J=2.44 meV, and an
anisotropy $D$ that is 0.093 meV above the spin reorientation transition at
40 K, and 0.126 meV below. For $H\parallel c$ the magnetic structures and
phase diagram have been determined, and reveal additional transitions below
8 K where the ferroelectrically displaced Ho$^{3+}$ ions are ordered
magnetically.
\end{abstract}

\maketitle





Magnetic ferroelectric materials are rare, and rarer still are systems where
these two disparate order parameters exhibit significant coupling
\cite{hill00}.  Such multiferroics have been of particular interest
recently both to understand the fundamental aspects of this coupling
\cite{sugie02,kimura03,hur04,lorenz04,lottermoser04}, and because of
the intriguing   
possibility of using these coupled order parameters in novel device
applications \cite{zheng04}. The hexagonal HoMnO$_{3}$ system of
particular interest here
is a prototype multiferroic where the Ho-O displacements give rise to a
ferroelectric moment (T$_{C}=875$K) along the c-axis, the Mn moments order
at 72 K, and the order parameters are naturally coupled through the Ho-Mn
exchange and anisotropy interactions. The magnetic system has the added
interest that the Mn moments occupy a fully frustrated triangular lattice. 
Our inelastic neutron scattering measurements of the Mn spin dynamics reveal
the planar nature of the spin system, and allow us to establish the basic
model for the magnetic interactions in the system and determine the
interaction parameters. Field-dependent neutron diffraction measurements
reveal the nature of the magnetic phase diagram, and in particular the Ho
involvement in the hysteretic magnetic transitions at low temperatures. 
Our results also suggest that the Ho coupling may be responsible for the
first-order spin-flop transition found at intermediate temperatures. 

\begin{figure}[tbp]
\includegraphics[width=8cm]{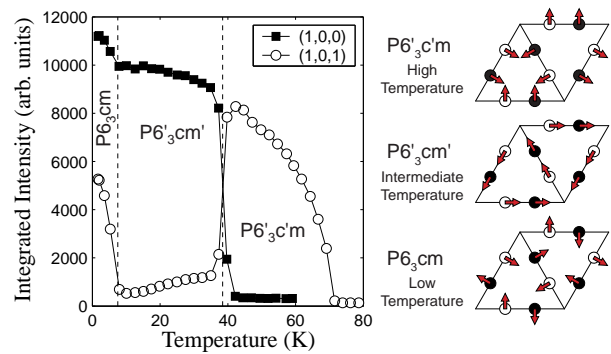}
\caption{(color online). Neutron diffraction measurements of the
(1,0,0) and (1,0,1) magnetic Bragg reflection integrated intensity.  Two
spin reorientation transitions (indicated by dashed lines) lead to changes in
the intensities of both peaks.  The Mn$^{3+}$ spin configurations
for these phases are shown in the schematics to the right, where
open circles indicate the position of Mn ions at $z=0$, filled
circles indicate Mn ions at $z=c/2$, and arrows indicate the direction
of the local magnetic moment.  Ho$^{3+}$ moments (not shown) order in
the low-temperature phase.}
\label{transitions}
\end{figure}

We have grown single crystals of HoMnO$_{3}$~using a traveling solvent
optical floating zone furnace, and the diffraction and inelastic neutron
measurements were performed at the NIST Center for Neutron Research on the
BT2 and BT9 thermal triple-axis instruments.
The magnetic structure of HoMnO$_{3}$~has been studied previously with
neutron powder diffraction \cite{koehler64,munoz01,lonkai02}, optical
second-harmonic generation (SHG)
\cite{fiebig00,fiebig00a,fiebig02,fiebig02b}, and magnetic and 
dielectric susceptibility \cite{sugie02}. The Mn$^{3+}$ ions are arranged in
a 2-dimensional (2D) triangular lattice, with successive layers offset from
each other. Fig.  \ref{transitions}  shows the integrated intensities of the
(1,0,0) and (1,0,1) (commensurate) magnetic Bragg peaks, where we see that
below $T_{N}\approx $ 72K the spin-2 Mn$^{3+}$ moments order
antiferromagnetically in a non-collinear 120-degree structure within each
plane. At $T_{SR}\approx $ 40 K a sharp spin reorientation transition is
observed associated with a change in the magnetic symmetry to  the $%
P6_{3}^{\prime }cm^{\prime }$~structure, as indicated in
Fig. \ref{transitions}. Below 8 K half of the Ho$^{3+}$ moments order
\cite{munoz01,lonkai02}, which results in a third zero-field spin
reorientation transition of the Mn spins. The three zero-field Mn spin 
structures are also shown in Fig. \ref{transitions}.

\begin{figure}[tbp]
\includegraphics[width=8.5cm]{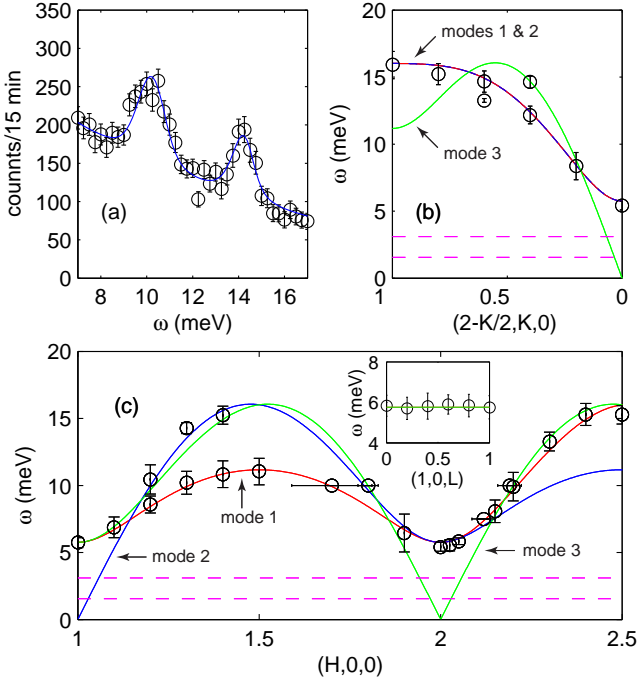}
\caption{(color online). Spin-wave dispersion at 20K. 
(a) Measurements of the spin waves at
Q=(1.3,0,0).  Data were taken on BT2 with horizontal collimations of
60'-40'-40'-80' and a fixed final energy of 14.7 meV. The line shows a fit
to the data using two modes convoluted with the resolution function. Peak
positions from both constant-energy and constant-Q scans are plotted for two
in-plane directions in reciprocal space in (b) and (c), along with the
three modes of 
the dispersion curve (solid lines) given by Eq. \ref{dispersion_eq} 
($J=2.44(7)$ meV and $D=0.126(25)$ meV). Dashed lines indicate two
(dispersionless) crystal field levels of Ho at 1.5(1) and 3.1(1) meV.
Inset: L dependence of the spin-wave spectrum, showing no discernable
dispersion.}
\label{Dispersion_20K}
\end{figure}

\begin{figure}[tbp]
\includegraphics[width=8cm]{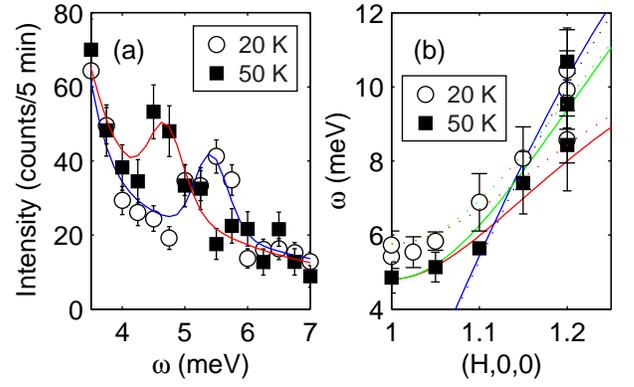}
\caption{(color online). (a) Inelastic neutron measurements at
Q=(2,0,0) at 20K and 50K.  Data were
taken on BT9 with horizontal collimations of 40'-40'-40'-40' and fixed
incident energy of 13.7 meV.  Lines show fits to the data.  The increase
in intensity at lower energies is primarily due to a crystal field
excitation at 3.1 meV.  (b)  Comparison of 20K and
50K dispersion data near
the zone center.  Data have been reduced to the first Brillouin zone.
Dotted lines show spin-wave dispersion curves for 20K data, as in
Fig. \ref{Dispersion_20K}.  Solid lines show fit of
Eq. \ref{dispersion_eq} to 50K data ($J=2.44$ meV is fixed and
$D=0.093(10)$ meV).} 
\label{gapcompare}
\end{figure}

Measurements of the spin-wave dispersion were performed on a 1.8 gram single
crystal in both the (H,K,0) and (H,0,L) scattering planes. An example of a 
constant-Q scan at 20K showing two distinct magnetic modes is shown in
Fig. \ref{Dispersion_20K}a.  Fits to individual 
scans were used to establish the peak positions. The dispersion relations at
20K, in the intermediate-temperature phase, are plotted in 
Fig. \ref{Dispersion_20K}b and \ref{Dispersion_20K}c along two
different directions in reciprocal space. Data for the spin-wave dispersion
along the (1,0,L) direction (inset Fig. \ref{Dispersion_20K}c) show
no discernible
dispersion in the out-of-plane direction, indicating primarily 2D spin
dynamics.  Data were also taken at 50K, in the high-temperature phase,
and a comparison of two constant-Q scans at the magnetic zone center
are shown in Fig. \ref{gapcompare}a.  The zone-center gap decreases
significantly, as shown in the comparison of the 20K and 50K
dispersion data in Fig. \ref{gapcompare}b.

The dispersion of the excitations in HoMnO$_{3}$~was calculated using a
linear spin-wave analysis. We start from the Hamiltonian 
\begin{eqnarray}
H=J\sum_{\langle i,j\rangle }\mathbf{S}_{i}\cdot \mathbf{S}%
_{j}+D\sum_{i}S_{i}^{z}S_{i}^{z},
\label{hamiltonian}
\end{eqnarray}
where $\langle i,j\rangle $ indicates the sum is over nearest-neighbor
in-plane pairs, $J$ is the primary antiferromagnetic exchange, and $D$ is
the anisotropy. Introducing three different flavors of bosons and using a
Holstein-Primakoff transformation, we derive a linearized Hamiltonian which
we diagonalize by mapping to the Hamiltonian of the quadratic quantum
mechanical oscillator, similarly to Ref \cite{harris92}.
We define the lattice Fourier sum as 
\begin{eqnarray}
z &=&\frac{1}{12}\left\{ e^{-2\pi i\left( H/3-K/3\right) }+e^{2\pi i\left(
2H/3+K/3\right) }\right.   \nonumber \\
&&\left. +e^{-2\pi i\left( H/3+2K/3\right) }\right\} .  \label{z_define}
\end{eqnarray}%
For simplicity, we now define 
\begin{eqnarray}
z_{1} &=&z+z^{\ast }  \nonumber \\
z_{2} &=&-(z+z^{\ast })/2+i(\sqrt{3}/2)(z-z^{\ast })  \nonumber \\
z_{3} &=&-(z+z^{\ast })/2-i(\sqrt{3}/2)(z-z^{\ast }).  \label{modes}
\end{eqnarray}%
We obtain three modes whose dispersion is given by 
\begin{eqnarray}
\omega _{i}=3SJ\sqrt{\left( 1+4z_{i}+2D/J\right) \left( 1-2z_{i}\right) },
\label{dispersion_eq}
\end{eqnarray}
where $i$ = 1, 2, or 3, and $S=2$ is the spin at the Mn$^{3+}$ ion. Equation 
\ref{dispersion_eq} was used to fit our data, with the results shown by the
curves in Figs. \ref{Dispersion_20K}b, \ref{Dispersion_20K}c, and 
\ref{gapcompare}b. This simple model provides a remarkably
good description of the spin dynamics. \ Quantitatively, we obtain $J=2.44(7)
$ meV and $D=0.126(25)$ at 20 K.  Data taken away from the zone center
showed little temperature dependence, while at the zone center there
was a significant difference between the 20K and 50K data, indicating
that the temperature dependence of the spin-wave spectrum comes
largely from changes in the anisotropy. For the 50K data, $J$ was
therefore kept fixed at 2.44 meV, yielding $D=0.093(10)$.  The
increase in the anisotropy $D$ at lower temperatures is signficant,
and suggests that this anisotropy, likely originating from the holmium,
drives the spin reorientation transition.


In the presence of a magnetic field $T_{SR}$ shifts to lower T and broadens,
and HoMnO$_{3}$~develops a reentrant $P6_{3}^{\prime }c^{\prime
}m$~phase. A sharp anomaly in the dielectric susceptibility at
$T_{SR}$ indicates a coupling between the magnetic and ferroelectric
order \cite{lorenz04}. SHG measurements as a
function of magnetic field also show the change in magnetic symmetry and the
reentrant phase as a function of magnetic field suggested by dielectric
susceptibility \cite{fiebig02}. More recent measurements show a marked
decrease in the strength of the SHG signal (which is due to the
magnetic ordering of the Mn$^{3+}$ ions) when an electric field is
applied \cite{lottermoser04}.

\begin{figure}[tbp]
\includegraphics[width=8cm]{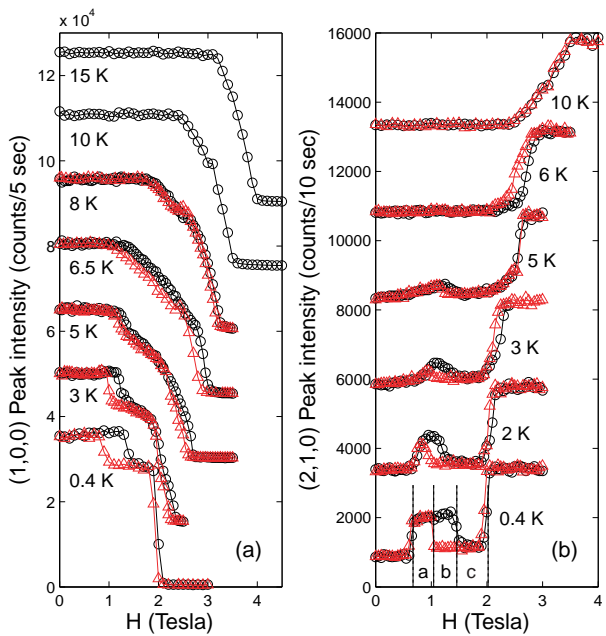}
\caption{(color online). (a) (1,0,0) magnetic Bragg peak intensity versus
magnetic field for different temperatures (offset by 15,000 counts for
clarity). (b) (2,1,0) magnetic Bragg peak
intensity versus magnetic field for different temperatures. Circles are data
taken with increasing field, while triangles are data taken with decreasing
field.  Dashed lines show 0.4K transitions, with two intermediate
phases $a$ and $c$ and hysteretic overlap in region $b$.}
\label{M_T_slice}
\end{figure}

\begin{figure}[tbp]
\includegraphics[width=8cm]{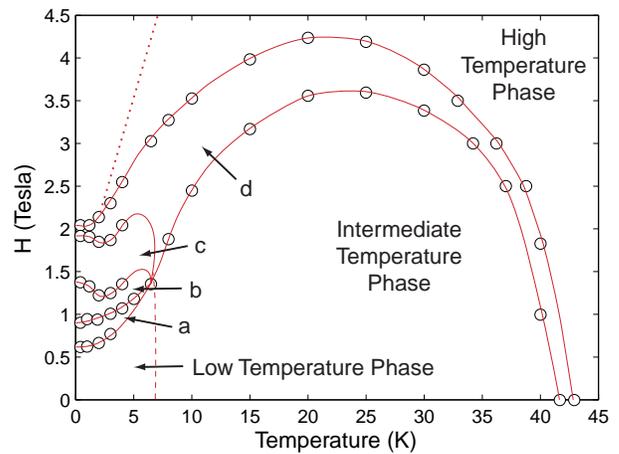}
\caption{(color online). Temperature and magnetic field phase diagram for
HoMnO$_{3}$~obtained from diffraction measurements of the (1,0,0) and
(2,1,0) Bragg reflections. Lines are guides to the eye.  Regions $a$
and $c$ are intermediate phases, with hysteretic overlap in region
$b$, as in Fig. \ref{M_T_slice}. Region $d$ is an intermediate region
where peak
intensities change continuously. Dashed line is approximate location of
low-temperature Ho$^{3+}$ ordering transition, and dotted line is
approximate transition observed by SHG and magnetic susceptibility
measurements \protect\cite{fiebig02,lorenz04b}. These transitions were not
observable from the (1,0,0) and (2,1,0) diffraction data. }
\label{phase_diagram}
\end{figure}

To investigate the magnetic field-dependent phase diagram, a 0.5 gram
sample was mounted in the (H,K,0) scattering plane 
inside a vertical-field 7 Tesla superconducting magnet with a Helium-3
insert. A magnetic field of up to 5 Tesla was applied along the $c$ axis,
and the (commensurate) (1,0,0) and (2,1,0) magnetic peaks were used to track
the evolution of the magnetic phases with magnetic field and temperature. In
the intermediate temperature phase, a sufficiently strong applied magnetic
field along the $c$ axis pushes HoMnO$_{3}$~into the high-temperature phase 
\cite{fiebig02,lorenz04}.
The boundary between the intermediate- and
high-temperature phases (shown in Fig. \ref{phase_diagram} as a
function of temperature and magnetic field) is in agreement with the
phase diagram established from dielectric susceptibility measurements
\cite{lorenz04,lorenz04b}.

At lower temperatures the transition becomes considerably more
complicated. Figure \ref{M_T_slice}a shows the peak intensity of the (1,0,0)
Bragg scattering as a function of magnetic field at several temperatures. At
higher temperatures, the intensity decreases approximately linearly with
increasing field over a well-defined transition region. At 0.4K, however,
there are two distinct, step-like transitions, with considerable hysteresis
in the lower-field transition, indicating the existence of a well-defined
intermediate phase between the low- and high-field phases. As the
temperature increases, these transitions become broader and the hysteresis
decreases, merging together into the single transition (without appreciable
hysteresis) observable at higher temperatures.

Figure \ref{M_T_slice}b shows equivalent measurements of the (2,1,0) Bragg
peak intensity. Again, at high temperatures, an approximately linear change
in scattering intensity over a well-defined transition region is observable.
At low temperatures, as for the (1,0,0) reflection, the situation becomes
more complicated. Instead of two, we find three transitions, with an
additional lower-field transition not observable in the (1,0,0)
data. Unlike the (1,0,0) reflection, the scattering intensity of the
(2,1,0) reflection does not change monotonically.

Figure \ref{phase_diagram} shows a phase diagram extracted from the (1,0,0)
and (2,1,0) diffraction data. Below ~4K
there are two distinct intermediate phases (labeled $a$ and $b$) between the
zero-field and high-field phases, with significant hysteretic overlap
between them. These data are not sensitive to the low-temperature spin
reorientation observable in the (2,1,0) diffraction data 
(Fig. \ref {transitions}), indicated roughly by the dashed line in 
Fig. \ref{phase_diagram}. 
It is interesting to note, however, that the intermediate
phases are only observable below the low-temperature spin reorientation
transition where Ho$^{3+}$ ordering becomes significant.


At higher temperatures, our phase diagram agrees well with previous
dielectric measurements \cite{lorenz04}. These SHG measurements did not
observe any intermediate phase at low temperatures \cite{fiebig02}, while
more recent magnetic susceptibility and heat capacity measurements have
identified one intermediate phase \cite{lorenz04b}. While we observe two
unambiguous, distinct intermediate phases at low temperatures, the combined
intermediate phases we observe agree qualitatively with the intermediate
phase recently observed \cite{lorenz04b}. Further measurements of
out-of-plane magnetic peaks will be necessary in order to determine the
exact magnetic structure of these intermediate phases. Recent SHG
measurements suggest that Ho$^{3+}$ magnetic moments can even be ordered
ferromagnetically with the application of an electric field 
\cite{lottermoser04}.
Neutron scattering reveals considerable complexity in the
magnetic field phase diagram for HoMnO$_{3}$, and extending these neutron
measurements to the electric field phase diagram may also prove similarly
revealing.

The out-of-plane magnetic interactions in HoMnO$_{3}$, both between
successive Mn$^{3+}$ layers and between Mn$^{3+}$ and Ho$^{3+}$, are almost
completely frustrated. It is therefore not surprising that the spin dynamics
in the intermediate and high temperature phases are two-dimensional in
nature, and we find that the spin-wave spectrum is well described by 
Eq. \ref{hamiltonian}. Ferroelectric lattice distortions relieve the
interplane frustrations in HoMnO$_3$ \cite{lottermoser04}, and may be
the driving force behind the 
spin-reorientation transitions of the Mn$^{3+}$ magnetic lattice.
The role of Ho in these reorientation
transitions may be very similar to the role of Nd in Nd$_2$CuO$_4$,
another layered transition-metal oxide in which frustrated interplanar
interactions enable weak, higher-order interactions to dictate the magnetic
structure.  Nd$_2$CuO$_4$ has a non-collinear structure
\cite{skanthakumar93} with Cu spin-reorientation transitions
\cite{skanthakumar89,endoh89} that are controlled by the
rare earth single-ion anisotropy \cite{sachidanandam97,petitgrand99}.
We suggest that it is a similar interaction in HoMnO$_3$ that controls
the spin-flop transition, and is the origin of the
magnetic-ferroelectric coupling. 

There has been considerable experimental
progress in recent years on mapping out the magnetic phase diagram of 
HoMnO$_{3}$
and discovering signatures of the interplay of magnetic and
ferroelectric order in various physical properties. However, a theoretical
understanding of these interactions is still lacking. Although a complete
theoretical description of HoMnO$_{3}$ will necessarily require
significantly more complexity than Eq. \ref{hamiltonian}, having
experimentally determined the dispersion relations and values for the
primary magnetic exchange constants is an important step toward being able
to model the interaction between ferroelectric and magnetic order in 
HoMnO$_{3}$.

\begin{acknowledgments}
Work at Rutgers was supported by NSF-MRSEC Grant No. DMR 00-80008. Work at
NIST was supported in part by the Binational Science Foundation, Grant No.
2000073. MK was supported by the National Science Foundation through
DMR-0306940.  Data reduction was performed with DAVE software,
supported by NSF Agreement No. DMR-0086210.
\end{acknowledgments}


\begin{thebibliography}{21}
\expandafter\ifx\csname natexlab\endcsname\relax\def\natexlab#1{#1}\fi
\expandafter\ifx\csname bibnamefont\endcsname\relax
  \def\bibnamefont#1{#1}\fi
\expandafter\ifx\csname bibfnamefont\endcsname\relax
  \def\bibfnamefont#1{#1}\fi
\expandafter\ifx\csname citenamefont\endcsname\relax
  \def\citenamefont#1{#1}\fi
\expandafter\ifx\csname url\endcsname\relax
  \def\url#1{\texttt{#1}}\fi
\expandafter\ifx\csname urlprefix\endcsname\relax\def\urlprefix{URL }\fi
\providecommand{\bibinfo}[2]{#2}
\providecommand{\eprint}[2][]{\url{#2}}

\bibitem[{\citenamefont{Hill}(2000)}]{hill00}
\bibinfo{author}{\bibfnamefont{N.~A.} \bibnamefont{Hill}}, \bibinfo{journal}{J.
  Phys. Chem. B} \textbf{\bibinfo{volume}{104}}, \bibinfo{pages}{6694}
  (\bibinfo{year}{2000}).

\bibitem[{\citenamefont{Sugie et~al.}(2002)\citenamefont{Sugie, Iwata, and
  Kohn}}]{sugie02}
\bibinfo{author}{\bibfnamefont{H.}~\bibnamefont{Sugie}},
  \bibinfo{author}{\bibfnamefont{N.}~\bibnamefont{Iwata}}, \bibnamefont{and}
  \bibinfo{author}{\bibfnamefont{K.}~\bibnamefont{Kohn}}, \bibinfo{journal}{J.
  Phys. Soc. Jpn.} \textbf{\bibinfo{volume}{71}}, \bibinfo{pages}{1558}
  (\bibinfo{year}{2002}).

\bibitem[{\citenamefont{Kimura et~al.}(2003)\citenamefont{Kimura, Goto,
  Shintani, Ishizaka, Arima, and Tokura}}]{kimura03}
\bibinfo{author}{\bibfnamefont{T.}~\bibnamefont{Kimura}},
  \bibinfo{author}{\bibfnamefont{T.}~\bibnamefont{Goto}},
  \bibinfo{author}{\bibfnamefont{M.}~\bibnamefont{Shintani}},
  \bibinfo{author}{\bibfnamefont{K.}~\bibnamefont{Ishizaka}},
  \bibinfo{author}{\bibfnamefont{T.}~\bibnamefont{Arima}}, \bibnamefont{and}
  \bibinfo{author}{\bibfnamefont{Y.}~\bibnamefont{Tokura}},
  \bibinfo{journal}{Nature} \textbf{\bibinfo{volume}{426}}, \bibinfo{pages}{55}
  (\bibinfo{year}{2003}).

\bibitem[{\citenamefont{Hur et~al.}(2004)\citenamefont{Hur, Park, Sharma, Ahn,
  Guha, and Cheong}}]{hur04}
\bibinfo{author}{\bibfnamefont{N.}~\bibnamefont{Hur}},
  \bibinfo{author}{\bibfnamefont{S.}~\bibnamefont{Park}},
  \bibinfo{author}{\bibfnamefont{P.~A.} \bibnamefont{Sharma}},
  \bibinfo{author}{\bibfnamefont{J.~S.} \bibnamefont{Ahn}},
  \bibinfo{author}{\bibfnamefont{S.}~\bibnamefont{Guha}}, \bibnamefont{and}
  \bibinfo{author}{\bibfnamefont{S.}~\bibnamefont{Cheong}},
  \bibinfo{journal}{Nature} \textbf{\bibinfo{volume}{429}},
  \bibinfo{pages}{392} (\bibinfo{year}{2004}).

\bibitem[{\citenamefont{Lorenz et~al.}(2004)\citenamefont{Lorenz, Litvinchuk,
  Gospodinov, and Chu}}]{lorenz04}
\bibinfo{author}{\bibfnamefont{B.}~\bibnamefont{Lorenz}},
  \bibinfo{author}{\bibfnamefont{A.~P.} \bibnamefont{Litvinchuk}},
  \bibinfo{author}{\bibfnamefont{M.~M.} \bibnamefont{Gospodinov}},
  \bibnamefont{and} \bibinfo{author}{\bibfnamefont{C.~W.} \bibnamefont{Chu}},
  \bibinfo{journal}{\prl} \textbf{\bibinfo{volume}{92}},
  \bibinfo{pages}{087204} (\bibinfo{year}{2004}).

\bibitem[{\citenamefont{Lottermoser et~al.}(2004)\citenamefont{Lottermoser,
  Lonkai, Amann, Hohlwein, Ihringer, and Fiebig}}]{lottermoser04}
\bibinfo{author}{\bibfnamefont{T.}~\bibnamefont{Lottermoser}},
  \bibinfo{author}{\bibfnamefont{T.}~\bibnamefont{Lonkai}},
  \bibinfo{author}{\bibfnamefont{U.}~\bibnamefont{Amann}},
  \bibinfo{author}{\bibfnamefont{D.}~\bibnamefont{Hohlwein}},
  \bibinfo{author}{\bibfnamefont{J.}~\bibnamefont{Ihringer}}, \bibnamefont{and}
  \bibinfo{author}{\bibfnamefont{M.}~\bibnamefont{Fiebig}},
  \bibinfo{journal}{Nature} \textbf{\bibinfo{volume}{430}},
  \bibinfo{pages}{541} (\bibinfo{year}{2004}).

\bibitem[{\citenamefont{{H. Zheng {\it et al.}}}(2004)}]{zheng04}
\bibinfo{author}{\bibnamefont{{H. Zheng {\it et al.}}}},
  \bibinfo{journal}{Science} \textbf{\bibinfo{volume}{303}},
  \bibinfo{pages}{661} (\bibinfo{year}{2004}).

\bibitem[{\citenamefont{Koehler et~al.}(1964)\citenamefont{Koehler, Yakel,
  Wollan, and Cable}}]{koehler64}
\bibinfo{author}{\bibfnamefont{W.~C.} \bibnamefont{Koehler}},
  \bibinfo{author}{\bibfnamefont{H.~L.} \bibnamefont{Yakel}},
  \bibinfo{author}{\bibfnamefont{E.~O.} \bibnamefont{Wollan}},
  \bibnamefont{and} \bibinfo{author}{\bibfnamefont{J.~W.} \bibnamefont{Cable}},
  \bibinfo{journal}{Phys. Lett.} \textbf{\bibinfo{volume}{9}},
  \bibinfo{pages}{93} (\bibinfo{year}{1964}).

\bibitem[{\citenamefont{{Mu\~noz} et~al.}(2001)\citenamefont{{Mu\~noz}, Alonso,
  {Mart\'inez-Lope}, Cas\'ais, Mart\'inez, and {Fern\'andez-D\'iaz}}}]{munoz01}
\bibinfo{author}{\bibfnamefont{A.}~\bibnamefont{{Mu\~noz}}},
  \bibinfo{author}{\bibfnamefont{J.~A.} \bibnamefont{Alonso}},
  \bibinfo{author}{\bibfnamefont{M.~J.} \bibnamefont{{Mart\'inez-Lope}}},
  \bibinfo{author}{\bibfnamefont{M.~T.} \bibnamefont{Cas\'ais}},
  \bibinfo{author}{\bibfnamefont{J.~L.} \bibnamefont{Mart\'inez}},
  \bibnamefont{and} \bibinfo{author}{\bibfnamefont{M.~T.}
  \bibnamefont{{Fern\'andez-D\'iaz}}}, \bibinfo{journal}{Chem. Mater.}
  \textbf{\bibinfo{volume}{13}}, \bibinfo{pages}{1497} (\bibinfo{year}{2001}).

\bibitem[{\citenamefont{Lonkai et~al.}(2002)\citenamefont{Lonkai, Hohlwein,
  Ihringer, and Prandl}}]{lonkai02}
\bibinfo{author}{\bibfnamefont{T.}~\bibnamefont{Lonkai}},
  \bibinfo{author}{\bibfnamefont{D.}~\bibnamefont{Hohlwein}},
  \bibinfo{author}{\bibfnamefont{J.}~\bibnamefont{Ihringer}}, \bibnamefont{and}
  \bibinfo{author}{\bibfnamefont{W.}~\bibnamefont{Prandl}},
  \bibinfo{journal}{Appl. Phys. A} \textbf{\bibinfo{volume}{74}},
  \bibinfo{pages}{S843} (\bibinfo{year}{2002}).

\bibitem[{\citenamefont{Fiebig et~al.}(2000{\natexlab{a}})\citenamefont{Fiebig,
  Fr\"ohlich, Kohn, Leute, Lottermoser, Pavlov, and Pisarev}}]{fiebig00}
\bibinfo{author}{\bibfnamefont{M.}~\bibnamefont{Fiebig}},
  \bibinfo{author}{\bibfnamefont{D.}~\bibnamefont{Fr\"ohlich}},
  \bibinfo{author}{\bibfnamefont{K.}~\bibnamefont{Kohn}},
  \bibinfo{author}{\bibfnamefont{S.}~\bibnamefont{Leute}},
  \bibinfo{author}{\bibfnamefont{T.}~\bibnamefont{Lottermoser}},
  \bibinfo{author}{\bibfnamefont{V.~V.} \bibnamefont{Pavlov}},
  \bibnamefont{and} \bibinfo{author}{\bibfnamefont{R.~V.}
  \bibnamefont{Pisarev}}, \bibinfo{journal}{\prl}
  \textbf{\bibinfo{volume}{84}}, \bibinfo{pages}{5620}
  (\bibinfo{year}{2000}{\natexlab{a}}).

\bibitem[{\citenamefont{Fiebig et~al.}(2000{\natexlab{b}})\citenamefont{Fiebig,
  Fr\"ohlich, Lottermoser, and Kohn}}]{fiebig00a}
\bibinfo{author}{\bibfnamefont{M.}~\bibnamefont{Fiebig}},
  \bibinfo{author}{\bibfnamefont{D.}~\bibnamefont{Fr\"ohlich}},
  \bibinfo{author}{\bibfnamefont{T.}~\bibnamefont{Lottermoser}},
  \bibnamefont{and} \bibinfo{author}{\bibfnamefont{K.}~\bibnamefont{Kohn}},
  \bibinfo{journal}{Appl. Phys. Lett.} \textbf{\bibinfo{volume}{77}},
  \bibinfo{pages}{4401} (\bibinfo{year}{2000}{\natexlab{b}}).

\bibitem[{\citenamefont{Fiebig et~al.}(2002{\natexlab{a}})\citenamefont{Fiebig,
  Degenhardt, and Pisarev}}]{fiebig02}
\bibinfo{author}{\bibfnamefont{M.}~\bibnamefont{Fiebig}},
  \bibinfo{author}{\bibfnamefont{C.}~\bibnamefont{Degenhardt}},
  \bibnamefont{and} \bibinfo{author}{\bibfnamefont{R.~V.}
  \bibnamefont{Pisarev}}, \bibinfo{journal}{J. Appl. Phys.}
  \textbf{\bibinfo{volume}{91}}, \bibinfo{pages}{8867}
  (\bibinfo{year}{2002}{\natexlab{a}}).

\bibitem[{\citenamefont{Fiebig et~al.}(2002{\natexlab{b}})\citenamefont{Fiebig,
  Fr\"ohlich, Lottermoser, and Maat}}]{fiebig02b}
\bibinfo{author}{\bibfnamefont{M.}~\bibnamefont{Fiebig}},
  \bibinfo{author}{\bibfnamefont{D.}~\bibnamefont{Fr\"ohlich}},
  \bibinfo{author}{\bibfnamefont{T.}~\bibnamefont{Lottermoser}},
  \bibnamefont{and} \bibinfo{author}{\bibfnamefont{M.}~\bibnamefont{Maat}},
  \bibinfo{journal}{\prb} \textbf{\bibinfo{volume}{66}},
  \bibinfo{pages}{144102} (\bibinfo{year}{2002}{\natexlab{b}}).

\bibitem[{\citenamefont{Harris et~al.}(1992)\citenamefont{Harris, Kallin, and
  Berlinsky}}]{harris92}
\bibinfo{author}{\bibfnamefont{A.~B.} \bibnamefont{Harris}},
  \bibinfo{author}{\bibfnamefont{C.}~\bibnamefont{Kallin}}, \bibnamefont{and}
  \bibinfo{author}{\bibfnamefont{A.~J.} \bibnamefont{Berlinsky}},
  \bibinfo{journal}{\prb} \textbf{\bibinfo{volume}{45}}, \bibinfo{pages}{2899}
  (\bibinfo{year}{1992}).

\bibitem[{lor()}]{lorenz04b}
\bibinfo{note}{B. Lorenz {\it et al.}, cond-mat/0408499}.

\bibitem[{\citenamefont{Skanthakumar et~al.}(1993)\citenamefont{Skanthakumar,
  Lynn, Peng, and Li}}]{skanthakumar93}
\bibinfo{author}{\bibfnamefont{S.}~\bibnamefont{Skanthakumar}},
  \bibinfo{author}{\bibfnamefont{J.~W.} \bibnamefont{Lynn}},
  \bibinfo{author}{\bibfnamefont{J.~L.} \bibnamefont{Peng}}, \bibnamefont{and}
  \bibinfo{author}{\bibfnamefont{Z.~Y.} \bibnamefont{Li}},
  \bibinfo{journal}{\prb} \textbf{\bibinfo{volume}{47}}, \bibinfo{pages}{R6173}
  (\bibinfo{year}{1993}).

\bibitem[{\citenamefont{Skanthakumar et~al.}(1989)\citenamefont{Skanthakumar,
  Zhang, Clinton, Li, Lynn, Fisk, and Cheong}}]{skanthakumar89}
\bibinfo{author}{\bibfnamefont{S.}~\bibnamefont{Skanthakumar}},
  \bibinfo{author}{\bibfnamefont{H.}~\bibnamefont{Zhang}},
  \bibinfo{author}{\bibfnamefont{T.~W.} \bibnamefont{Clinton}},
  \bibinfo{author}{\bibfnamefont{W.}~\bibnamefont{Li}},
  \bibinfo{author}{\bibfnamefont{J.~W.} \bibnamefont{Lynn}},
  \bibinfo{author}{\bibfnamefont{Z.}~\bibnamefont{Fisk}}, \bibnamefont{and}
  \bibinfo{author}{\bibfnamefont{S.}~\bibnamefont{Cheong}},
  \bibinfo{journal}{Physica C} \textbf{\bibinfo{volume}{160}},
  \bibinfo{pages}{124} (\bibinfo{year}{1989}).

\bibitem[{\citenamefont{Endoh et~al.}(1989)\citenamefont{Endoh, Matsuda,
  Yamada, Kakurai, Hidaka, Shirane, and Birgeneau}}]{endoh89}
\bibinfo{author}{\bibfnamefont{Y.}~\bibnamefont{Endoh}},
  \bibinfo{author}{\bibfnamefont{M.}~\bibnamefont{Matsuda}},
  \bibinfo{author}{\bibfnamefont{K.}~\bibnamefont{Yamada}},
  \bibinfo{author}{\bibfnamefont{K.}~\bibnamefont{Kakurai}},
  \bibinfo{author}{\bibfnamefont{Y.}~\bibnamefont{Hidaka}},
  \bibinfo{author}{\bibfnamefont{G.}~\bibnamefont{Shirane}}, \bibnamefont{and}
  \bibinfo{author}{\bibfnamefont{R.~J.} \bibnamefont{Birgeneau}},
  \bibinfo{journal}{\prb} \textbf{\bibinfo{volume}{40}}, \bibinfo{pages}{7023}
  (\bibinfo{year}{1989}).

\bibitem[{\citenamefont{Sachidanandam et~al.}(1997)\citenamefont{Sachidanandam,
  Yildirim, Harris, Aharony, and Entin-Wohlman}}]{sachidanandam97}
\bibinfo{author}{\bibfnamefont{R.}~\bibnamefont{Sachidanandam}},
  \bibinfo{author}{\bibfnamefont{T.}~\bibnamefont{Yildirim}},
  \bibinfo{author}{\bibfnamefont{A.~B.} \bibnamefont{Harris}},
  \bibinfo{author}{\bibfnamefont{A.}~\bibnamefont{Aharony}}, \bibnamefont{and}
  \bibinfo{author}{\bibfnamefont{O.}~\bibnamefont{Entin-Wohlman}},
  \bibinfo{journal}{\prb} \textbf{\bibinfo{volume}{56}}, \bibinfo{pages}{260}
  (\bibinfo{year}{1997}).

\bibitem[{\citenamefont{Petitgrand et~al.}(1999)\citenamefont{Petitgrand,
  Maleyev, Bourges, and Ivanov}}]{petitgrand99}
\bibinfo{author}{\bibfnamefont{D.}~\bibnamefont{Petitgrand}},
  \bibinfo{author}{\bibfnamefont{S.~V.} \bibnamefont{Maleyev}},
  \bibinfo{author}{\bibfnamefont{P.}~\bibnamefont{Bourges}}, \bibnamefont{and}
  \bibinfo{author}{\bibfnamefont{A.~S.} \bibnamefont{Ivanov}},
  \bibinfo{journal}{\prb} \textbf{\bibinfo{volume}{59}}, \bibinfo{pages}{1079}
  (\bibinfo{year}{1999}).

\end{thebibliography}

\end{document}